# PREPRINT[1]: Practical Knowledge Management Tool Use in a Software Consulting Company

Torgeir Dingsøyr[1], Hans Karim Djarraya[2], Emil Røyrvik[3]

[1]SINTEF Information and Communiation Technology,

7465 Trondheim, Norway

Telephone: +47 73 59 29 79, Telefax: +47 73 59 29 77

torgeir.dingsoyr@sintef.no

[2]Computas AS

Box 482, N-1327 Lysaker, Norway

Telephone: +47 67 83 10 22, Telefax: +47 67 83 10 01

hkd@computas.com

[3]SINTEF Technology and Society,

7465 Trondheim, Norway

Telephone: +47 73 55 05 24, Telefax: +47 73 59 12 99

emil.royrvik@sintef.no

---

Tools for managing technical skills are used in many companies, but there has been little discussion about how such tools are used in practice. We report here on different types of actual usage in a medium-size software consulting company. We expected such tools to be used for allocating resources to new projects and for searching for competence to solve problems, but also observed two other types of usage: identifying new project opportunities, and skills upgrading. This multitude of uses support learning practices and motivates tool use both at individual and company levels, which is crucial to support organizational learning.

**Managing Technical Skills**

Software engineering is a knowledge-intensive task, where it is obviously critical to have skilled employees. Knowledge management [1] has gained much attention amongst practitioners and researchers in software engineering [2, 3]. There are two basic strategies in ensuring that employees are skilled [4]: Either you focus on codifying relevant knowledge, or you count on communication between people who have relevant knowledge – what is called "personalization". Kankanhalli et. al [5] gives examples of groups of companies who have chosen different strategies, and discusses the role of IT for supporting the strategies. In this article, we seek to further discuss one type of IT support for a personalization strategy, namely the use of tools to manage and identify technical skills of employees.

Such systems are sometimes referred to as expert directories, people-finder systems or company-internal yellow pages.

We can divide skills in two broad groups – *technical skills:* Knowledge about technology issues, and more *soft skills:* Competencies of a more personal and social flavour, like organising and handling complexities in project work, enabling people to contribute with their resources, and customer communication.

It is of major importance to get the right people with the right soft and technical skills to work on a software development project. Many companies have developed knowledge management tools to assist them in the tasks of managing *technical skills*, by surveying what kind of knowledge people have, and make an index of it. The process of surveying and indexing and making this type of information available, we will refer to as *skills management,* and we will focus on technical skills in the following.

There are many software tools for managing skills – for example, companies that offer jobs on the Internet usually have a database where you as a job-seeker can store your competence profile. The contents of such tools can be: "Knowledge profiles, skill profiles and personal characteristic profiles that define subjective assessments of the knowledge, skills, and personal traits required for the different work-roles" [6].

In order to have such a working system, a company needs to select a set of skills that they are interested in, have a system for evaluating the employees, and make this information available to different user groups.

We wanted to know more about how tools for managing skills are *used* in a specific organization. What purposes do such tools serve, and do they satisfy needs other than the expected use in resource allocation? In order to examine these questions, we interviewed 14 developers, managers and project managers in an ethnographic study at Computas - a consultancy company that develops knowledge-based software with about 150 employees. The company has no traditional departments, but is organized in projects and a set of processes, where knowledge management is considered to be one important process.

We now go on to describe their *skills management* tool, and what different types of usage we found. Note that we do not look at the technical implementation of skills management systems, but refer readers with such interests to other literature [7, 8]. A more thorough description of our findings is available in [9], and studies of usage of other knowledge management tools in the same company can be found in [10] and [11].

**The Skills Manager Tool**

The skills manager is a part of the Intranet at Computas, and every employee has access to it. You can select a skill from a taxonomy of around 250 different technical skills, related to the core competence of the company. When you select a skill, you can find which of seven skill levels people have, from "expert" to "irrelevant". In addition to indicating their skill level, people also indicate which level they want to have in the future, see Figure 1. When viewing skills, details are shown in black if people are on the level they would like to be on, red if this is a topic they do not wish to work on in future, or green if they want to develop their skills in this direction.

Employees are prompted in the front page of the Intranet to evaluate themselves when new skills are introduced in the tool. They are also told to update the information when they have completed a project. Anyone can suggest new skills to the tool, which will be included by the manager of the "competence center" process.

**A Variety of Usages**

When we think of usage of Skills Management tools, we would normally think of resource allocation and searching for experts to solve problems. But when we interviewed people at Computas, we found four ways of using the skills management tool, and we found issues that broadend our understanding of the usage of the tool.

**Resource allocation**

Concerning the classical usage of skills management tools, resource allocation, we found evidence of such practice. As one new employee said: "Contrary to a lot of other companies that use such a system, here at Computas we *really use* the system for resource planning." Another comment is in the same vein: "I think that the skills manager is a useful tool, but a tool that still has got a lot of potential when it comes to practical use. Those responsible for resource-management already use the tool a lot in their daily work." A third Computas employee also comments on the skills manager as an important tool for resource allocation, and indicates the dynamics whereby new skills are added to the tool: "The tools I use the most I think are […] the competence-block-manager [another part of Computas' knowledge management system, used for organizing internal learning courses] and the skills manager. Definitely! I'm responsible for the content in many databases, and partly the skills management base. And the skills manager is a tool that is very important […] for the resource allocation process […] Therefore, many employees come up with suggestions on new content, new elements, in the skills database."

**Searching for competence to solve problems**

Often developers need to acquire knowledge on some issue where they have little experience themselves. One developer describes a "short term" usage in solving problems: "Of course, when I wonder if there is anyone who can help me with something, I look up in the skills management system to see if anyone has the knowledge that I need". When you get a list of people with a certain competence, you can e-mail one or all of them. Or you can just print a list of people and ask them yourself, as another developer is usually doing. Of course, this depends on people rating themselves in an honest way. According to one developer, "Some overrate themselves and other underrate themselves strongly".

The last problem, can be termed "resistance to be known as an expert" [12]. If a developer indicates expertise in one area, she risks being allocated to that kind of project in the future instead of more intellectually challenging ones. The possibility of indicating your future wanted skill level is a means to prevent this resistance in this company. Thus, the tool signals not only what an employee is skilled at, but also topics he wants to work more with in the future.

Another developer is critical to the categories of competence in the skills management tool: "when it comes to more detailed things, like who can in fact write a computer program, and who can find a solution – you do not find that there".

In this usage domain we found both "short term" and "long term" types of usage. The former is where people use the tool to find others in the company that know something about a specific problem at hand. The latter is a usage where people over time increase their overall insight into what core competencies exist in the company.

When we look at the long-term usage, one developer says he finds a group of people that knows something about a subject in the skills management tool, and asks them questions by e-mail. But usually only a few people bother to answer: "if you have asked questions about SQL to ten gurus, and it is always the same two that answer, you start to go directly to them and talk; you learn after a while who it is any use to attempt to get information from".

**Finding projects and external marketing**

Another usage of the tool is for the sales department. One manager said that "Even sales can use it [the skills management system], to find new directions to go in". That is, to find what types of projects that suit the company well; combining strategic and competence development needs. Another usage could be to use the tool as a testimony of the organization's skill level. For Computas, as a company focusing on knowledge-based solutions, the Skills Manager is also a way to show that the company is taking its own medicine.

**Skills upgrading**

At Computas, people are allocated to projects on the basis of defined technical skills in the Skills Manager. In this way, people position themselves for future projects by indicating what knowledge they want to develop as a part of their career plan. And it is "natural to ask for an update on competencies when a project is finished". One employee sees the Skills manager in terms of intellectual capital: "[We can] say that we have that many man months with C++ competence, or Java, and we see that there is an increase in this competence, and then we can evaluate that." And by stating what they want to learn about in the future, people can develop their competence by working on relevant projects.

**Implications**
The study shows how a skills manager tool can be practically utilized in a variety of uses. Some implications of the findings are as follows:

- The design and user interface of the tool is simple and flexible, as a web-based system with update requests on the Intranet. It invites the people with the knowledge and the needs to suggest topics and contents to be rated. This leads to appropriate levels of detail in the competency areas. The topics should not be too detailed, and not too general.

- A broader span of usage functionality spurs utility. An important factor here is that the tool is advantageous at the individual level (employees get recognition for their skills, and sell themselves for future project assignments), at the project level (resource allocation) and at the process level (strategic management of core competencies).

- The double individual/company utility of the tool represents a wide scope of applicability, and is apparent to users. This motivates them to update skills information regularly. Updating is a common challenge for knowledge management systems.

- The tool is not designed for employees to get direct access to all relevant information on different topics. The skills manager is in tune with the personalization school of knowledge management, which seeks to connect people and problems, tasks and occasions. This strategy enables sharing of more subtle forms of knowledge, for example tacit knowledge.
- The tight integration of the tool with the daily, practical project work for most of the roles in the company ensures its use, usefulness and success.

**Conclusions**

We interviewed people in a software consultancy company about how they are using a *skills management* tool - a part of the company's knowledge management system. We expected the tool to be used for resource allocation and for short-term problem-solving. We found that the tool use in problem-solving also has a long-term effect in letting employees know who to ask next time. Further, the *skills management* tool is used for identifying new project opportunities and to support skills upgrading. The tool supports learning practices and motivates use at both an individual and company (projects and processes) level. This double capability enables integration and knowledge exchanges both vertically (between organisational levels) and horizontally (between individuals and projects). The skills manager thus supports processes that drive *organizational* learning – an increasingly recognised factor for competitive advantage.

Some employees are critical of how people evaluate their skills, others questioned the level of detail on the available skills, and yet others felt that information on more soft skills was lacking.

But all in all, it seems that the usage of the tool is very much implanted in the daily work of the organization, and supports a multitude of functions.

# References

[1] Meinolf Dierkes, Ariane Berthoin Antal, John Child, and Ikujiro Nonaka, *Handbook of Organizational Learning and Knowledge*: Oxford University Press, 2001, ISBN 0-19-829583-9.
[2] Mikael Lindvall and Ioana Rus, "Knowledge Management in Software Engineering," *IEEE Software*, no. 3, vol. 19, pp. 26 - 38, 2002.
[3] A. Aurum, R. Jeffery, C. Wohlin, and M. Handzic, *Managing Software Engineering Knowledge*. Berlin: Springer Verlag, 2003,
[4] Morten T. Hansen, Nitin Nohria, and Thomas Tierney, "What is your strategy for managing knowledge?," *Harvard Business Review*, no. 2, vol. 77, pp. 106 - 116, 1999.
[5] Atreyi Kankanhalli, Fransiska Tanudidjaja, Juliana Sutanto, and and Bernard C. Y. Tan, "The Role of IT in Successful Knowledge Management Initiatives," *Communications of the ACM*, no. 46, vol. 9, 2003.
[6] Karl M. Wiig, *Knowledge Management Methods*: Schema Press, 1995, ISBN 0-9638925-2-5.
[7] M Maybury, R M A, "Awareness of organizational expertise," *International journal of human-computer interaction*, no. 2, vol. 14, pp. 199-217, 2002.
[8] Irma Becerra-Fernandez, "The role of artificial intelligence technologies in the implementation of People-Finder knowledge management systems," *Knowledge-Based Systems*, no. 5, vol. 13, pp. 315-320, 2000.
[9] Torgeir Dingsøyr and Emil Røyrvik, "Skills Management as Knowledge Technology in a Software Consultancy Company," in *Proceedings of the Learning Software Organizations Workshop*, *Lecture Notes in Computer Science*, vol. 2176, K.-D. Althoff, R. L. Feldmann, and W. Müller, Eds. Kaiserslautern, Germany: Springer Verlag, 2001, pp. 96-107.
[10] Torgeir Dingsøyr, "Knowledge Management in Medium-Sized Software Consulting Companies," doctoral thesis, *Department of Computer and Information Science*, Norwegian University of Science and Technology, Trondheim, 2002, pp. 206, ISBN 82-7477-107-9.
[11] Gunnar John Coll, Steinar Carlsen, and Åsmund Mæhle, "Activity-centred Knowledge Support," in *Living knowledge: The dynamics of professional service work*, A. Carlsen, R. Klev, and G. von Krogh, Eds. Houndmills: Palgrave Macmillan, 2004, pp. 204 - 221.
[12] Kevin C. Desouza, "Barriers to Effective Use of Knoweldge Management in Software Engineering," *Communications of the ACM*, no. 1, vol. 46, 2003.